\documentclass[twocolumn,nofootinbib,superscriptaddress]{revtex4-1}
\usepackage{latexsym,epsfig,amssymb, amsmath, nicefrac}

\usepackage{color}
\usepackage[makeroom]{cancel}
  \usepackage{latexsym}
  \usepackage{epsf}
  \usepackage{amssymb}
  \usepackage{graphicx}
  \usepackage{amsmath}
  \usepackage{amsmath,amssymb,amsthm}
  \usepackage{verbatim}
  \usepackage{hyperref}

\def\tb0{\tilde{\beta}_0}
{\def\b0{\beta_0}

\def\bi{\begin{itemize}}
\def\ei{\end{itemize}}
\def\be{\begin{equation}}
\def\ee{\end{equation}}
\newcommand{\bea}{\begin{eqnarray}}
\newcommand{\eea}{\end{eqnarray}}

\def\a{\alpha}
\def\b{\beta}

\def\vf{\varphi}

\def\m{\mu}
\def\n{\nu}

\def\o{\omega}

\def\Poincare{Poincar\'e~}
\def\Mpl{M_{\rm Pl}}
\def\nn{\nonumber}

\def\empty{}
\def\emptyy{}

\begin{document}

\vspace{1cm}

\title{Universality Classes of Scale Invariant Inflation}

\author{Mehmet Ozkan}
\email{m.ozkan@rug.nl}
\author{Diederik Roest}
\email{d.roest@rug.nl}
\affiliation{Van Swinderen Institute for Particle Physics and Gravity, University of Groningen, \\ Nijenborgh 4, 9747 AG Groningen, The Netherlands}

\begin{abstract}
We investigate the inflationary implications of extensions of \Poincare symmetry. The simplest constructions with local scale invariance lead to universal predictions: the spectral index is $n_s = 1-2/N$, in excellent agreement with Planck data, while the tensor-to-scalar ratio is determined by a free parameter to $r = 12 \alpha / N^2$. For the special value $\alpha=1$ one finds symmetry enhancement to the full conformal group. We show that these findings hold both for two-derivative scalar-tensor theories as well as higher-derivative gravity. Therefore scale invariance underlies a promising set of inflationary models.
\end{abstract}	

\maketitle

\section{Introduction}

What is the fundamental symmetry that underlies the laws of Nature, and how can we test for it -- are there observational indications for symmetries beyond Poincar\'{e}? In this letter we will argue that the most recent measurements of the temperature fluctuations of the cosmic microwave background (CMB) suggest an extension of the Poincar\'e symmetry with a dilatation or scaling generator. Moreover, a future detection of primordial gravitational waves in the B-mode polarization of the CMB could point towards a further extension to the conformal group.

\begin{figure}
\centering
\includegraphics[width=.48\textwidth]{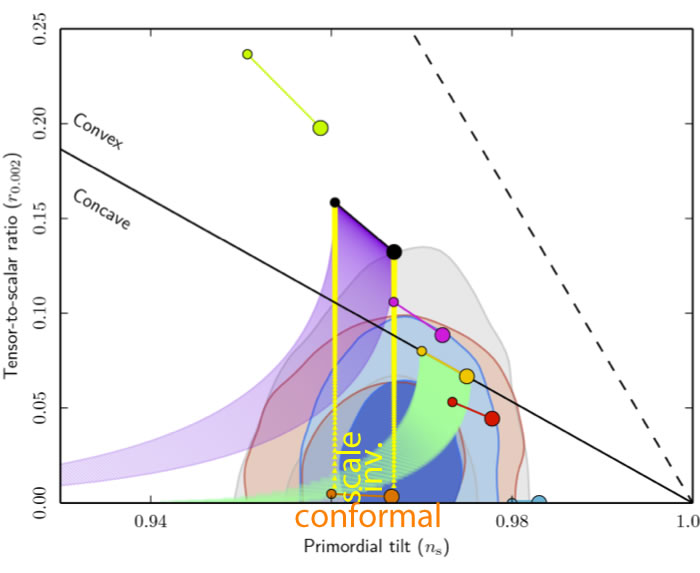}
\caption{\it The Planck 2015 one- and two-sigma contour plots for $n_s$ and $r$, with the predictions of a number of models superimposed \cite{Planck2015}: the yellow triangles stem from $\alpha$-attractors with $V = V_0 \tanh(\varphi)^2$, while the orange dots include the Starobinsky model as well as conformal inflation \cite{Starobinsky:1980, conformal-inflation}. We have included the underlying symmetry in the corresponding color. \label{fig1}}
\vspace{-0.5cm}
\end{figure}

Due to impressively accurate measurements of CMB, in particular the TT power spectum and TE crosscorrelation, the Planck satellite has put tight constraints on the scale dependence of the primordial scalar power spectrum. This is encoded in the spectral index $n_s = 1 - d \log \Delta^2_R / dN$, where $N$ is the number of e-folds from CMB horizon crossing to the end of inflation. The Planck 2015 data gives \cite{Planck2015}
 \begin{align}
   n_s = 0.968 \pm 0.006 \,.
 \end{align}
In addition it poses constraints on the power spectrum of primordial tensor fluctuations: the tensor-to-scalar ratio $r$ is constrained by the $2 \sigma$ upper limit $r<0.11$, as can be seen in Fig.~\ref{fig1}.

Similar results from the Planck 2013 data release have spurred an exciting development in inflationary model-building. In particular, it has been noted that the measured value is perfectly consistent with $n_s = 1 - 2/N$ for $N= 50 - 60$. Interestingly, this behaviour has been found to arise in a wide variety of inflationary models based on (super-)symmetry, non-minimal couplings and/or non-canonical kinetric terms \cite{Ellis:2013, Kallosh:2013wya, conformal-inflation, Ferrara:2013, Kallosh:2013tua, Kallosh:2013yoa, Kehagias:2013}. An important trait that many of these models share is an almost complete insensitivity of the inflationary predictions to the details of the models; it is this robustness that gives rise to the notion of cosmological attractors \cite{Galante:2014}. Amongst the predictions of these models are $1/N^2$ scaling of both the tensor-to-scalar ratio as well as the running of the scalar index. This is a specific case of the possible $N$-dependences identified in  \cite{Mukhanov:2013, Roest:2013}. 

In this letter we will provide a formal demonstration in  simple constructions, both at the two-  as well as higher-derivative level, that an additional symmetry underlies these attractor constructions. This has already been pointed out at the two-derivative level for the special case $r = 12 / N^2$, in which case there is a conformal symmetry \cite{conformal-inflation}. We will demonstrate that there is a one-parameter family of universality classes when only imposing a scale symmetry, where the free parameter generalizes the possible values of $r$. Both the conformal as well as scale invariant predictions are superimposed in Fig.~\ref{fig1}.

\section{Gravity as a gauge theory}

In this section we will outline the construction of gravity as a \Poincare gauge theory as well as two extensions, basen on the scale invariant and the conformal symmetries. 

The \Poincare group has ten generators, consisting of four translations $P_a$ as well as six Lorentz boosts and rotations $M_{ab}$. This symmetry can be gauged by the introduction of the corresponding gauge fields $e_{\mu}{}^a$ and $\omega_{\mu}{}^{ab}$. Their field strengths are given by
\begin{align}
  R_{\m\n}{}^a (P) =& 2 \partial_{[\m} e_{\n]}{}^a + 2 \o_{[\m}{}^{ab} e_{\n]b} \,, \notag \\
R_{\m\n}{}^{ab}(M) =& 2\partial_{[\mu}\omega_{\nu]}{}^{ab}+2\,\omega_{[\mu}{}^{ac}\,\omega_{\nu]c}{}^{b} \,.
 \end{align}
In order to obtain general relativity, with only the metric as independent, propagating field, one has to impose the curvature condition $R_{\mu \nu}{}^a (P) = 0$. This allows one to identify the gauge field $\omega_{\mu}{}^{ab}$ as the spin connection, i.e.~the Levi-Civita connection of the Vielbein $e_{\mu}{}^a$. At this point one can construct the most general diffeomorphism invariant theory, with arbitrary couplings for the metric field as well as any scalar fields that one would introduce: the \Poincare symmetry poses no further constraints apart from general coordinate invariance.

Augmenting the \Poincare symmetry with a dilatation or scaling generator $D$ (under which only the Abelian $P_a$ can have a non-trivial weight), we need to add an additional gauge field $b_\mu$. The curvatures now read
   \begin{align}
R_{\m\n}{}^a (P) =& 2 \partial_{[\m} e_{\n]}{}^a + 2 \o_{[\m}{}^{ab} e_{\n]b} + 2 b_{[\m} e_{\n]}{}^a \,,\notag \\
R_{\m\n}{}^{ab}(M) =& 2\partial_{[\mu}\emptyy\omega_{\nu]}{}^{ab}+2\,\emptyy\omega_{[\mu}{}^{ac}\,\emptyy\omega_{\nu]c}{}^{b}  \,,\notag\\
R_{\m\n} (D) =& 2 \partial_{[\m} b_{\n]} \,.
\end{align}
Again we impose the curvature condition $R_{\m\n}{}^a (P) = 0$, allowing us to solve for the would-be spin connection in terms of the Levi-Civita connection and a term with the additional gauge field:
 \begin{align}
  {\omega}_\mu{}^{ab} & = & 2e^{\nu[a}\partial_{[\mu}e_{\nu]}{}^{b]}-e^{\nu[a}e^{b]\sigma}e_{\mu c}\,\partial_{\nu}e_{\sigma}{}^{c}+2e_{\mu}{}^{[a}b^{b]} \,. 
 \end{align} 
   The Bianchi identity implies
 \begin{align}
   e_{[\mu}{}^a R_{\nu \rho]} (D) = R_{\mu \nu \rho}{}^a (M) \,.
 \end{align}
The field strength of the gauge field $b_\mu$ therefore does not necessarily vanish, and one could add a kinetic term for this field; however, we will not do so in order to have only the metric with propagating degrees of freedom.

The additional dilatation symmetry poses constraints on the possible dynamics that one can introduce. The Einstein-Hilbert term, for instance, can be obtained from a contraction of $R_{\m\n}{}^{ab}(M)$; in this combination, however,  the gauge field $b_\mu$ is also present \cite{Zhang:1990nm}:
 \begin{align}
  R(M) = R - 6 \nabla_\mu b^\mu - 6 b_\mu b^\mu \,.
 \end{align}
Similarly, the kinetic term for a scalar field $\phi$ with a non-vanishing scaling weight (which we take to be one) will also involve the same gauge field, since
 \begin{align}
     \Box^s \phi = (\partial^a - 2 b^a + \o_b{}^{ba}) (\partial_a - b_a) \phi \,,  
  \end{align}
where $\Box^s$ denotes the scale invariant d'Alambertian. As we will see, the gauge field $b_\mu$ that appears in these covariant quantities will play an important role in what follows.

Finally, we consider the extension to the full conformal symmetry $SO(4,2)$. In addition to \Poincare and scaling, this group has four special conformal generators $K_a$, with corresponding gauge fields $f_{\mu}{}^a$. The curvatures of the conformal group are (see e.g.~\cite{Freedman:2012})
\bea
R_{\m\n}{}^a (P) &=& 2 \partial_{[\m} e_{\n]}{}^a + 2 \empty\o_{[\m}{}^{ab} e_{\n]b} + 2 b_{[\m} e_{\n]}{}^a \,,\nn\\
R_{\m\n}{}^{ab}(M) &=& 2\partial_{[\mu}\empty\omega_{\nu]}{}^{ab}+2\,\empty\omega_{[\mu}{}^{ac}\,\empty\omega_{\nu]c}{}^{b}+8f_{[\mu}{}^{[a}e_{\nu]}{}^{b]} \,,\nn\\
R_{\m\n} (D) &=& 2 \partial_{[\m} b_{\n]} - 4 f_{[\m}{}^a e_{\n]a} \,,\nn\\
R_{\m\n}{}^{a} (K) &=& 2\partial_{[\m} f_{\n]}{}^a + 2 \empty\o_{[\m}{}^{ab} f_{\n]b} - 2 b_{[\m} f_{\n]}{}^a \,.
\eea  
In order to eliminate the gauge fields $\omega_{\mu}{}^{ab}$ and $f_\mu{}^a$ as independent fields, we now impose the curvature conditions $R_{\mu \nu}{}^a (P) = 0$ and $R_{\mu}{}^a (M) = 0$.  In this case the Bianchi identities imply that $R_{\mu \nu} (D) = 0$ as well as $R_{[\mu \nu \rho]}{}^a (M) = 0$. Therefore, in the conformal theory the $b_\mu$ field is pure gauge, and the only propagating field is the metric.

As expected, the conformal symmetry poses yet stronger constraints on the possible dynamics. Due to the curvature constraints, $R(M)$ vanishes in this case. Instead, the Einstein-Hilbert action appears in the d'Alembertian of a scalar field $\phi$ of weight one:
 \begin{align}
  \Box^c \phi = \Box \phi - \tfrac16 \phi R \,.
 \end{align}
This combination transforms covariantly with scaling weight three. Note that the gauge field $b_\mu$ drops out of this combination; this is due to the fact that the only independent gauge field that transforms under the special conformal symmetry is $b_\m$. 

The fact that $b_\m$ drops out in the conformal set-up, with a fixed ratio between the coefficients of the Einstein-Hilbert term and the d'Alembertian of the scalar field $\phi$, has a dramatic consequence: the conserved current that is associated with the dilatation symmetry vanishes \cite{Jackiw:2014, Campigotto:2014}. For the scale invariant set-ups, there is a non-zero conserved current for dilatations, which vanishes when the scale invariance is enhanced to conformal invariance \cite{Oda:2013uca}. At the same time, the trace of the energy-momentum tensor vanishes in the conformal but not in the scale invariant case. 

There are two situations  in which scale invariance implies the full conformal symmetry. First of all, in the off-shell supersymmetric extension of the locally scale-invariant gravity models, one needs enhancement to the full conformal symmetry in order to match the number of fermionic and bosonic degrees of freedom. Secondly, in two dimensions, it was proven that  scale invariance together with unitarity implies conformal invariance \cite{Zamolodchikov:1986gt, Polchinski:1987dy}. Therefore, it is intriguing to consider two dimensional scale invariant gravity to see whether the theory also exhibits conformal invariance. Turning our attention to $d=2$, we find that the definition of conformal and scale invariant d'Alambertians coincide, and neither of them include an Einstein-Hilbert term. Indeed, the scale invariant completion of the Einstein-Hilbert term contains $b_\m$ as a total derivative, which can be simply be dropped out in the action.

\section{Scalar-tensor theory}

What are the implications of the constrained dynamics on the possible inflationary models?  In order to address this question, we add a propagating scalar degree of freedom. Due to the scale symmetry, this requires the introduction of a pair of scalar fields with weight one, that will be referred to as the dilaton or conformon $\chi$ and the inflaton $\phi$. 

Following \cite{conformal-inflation} we will write the kinetic terms for these in an $SO(1,1)$ invariant combination. This symmetry will correspond to the non-compact shift symmetry of the inflaton, which will only be broken by the scalar potential. This leads to
\begin{align}
\mathcal{L} = & \frac{1}{12} (\chi^2 - \phi^2) R(M) - \frac12 \a (\chi \Box^s \chi - \phi \Box^s \phi ) + \notag \\
&  -  \frac{1}{36} F\Big(\frac{\phi}{\chi}\Big) (\phi^2 - \chi^2)^2 \,.
\label{2scalaractionS}
\end{align}
Since we have an independent curvature and kinetic term, we have included a relative coefficient $\alpha$ between these. Moreover, all terms are scale invariant and only the function $F$ breaks the $SO(1,1)$ symmetry by introducing a non-trivial profile to the scalar potential\footnote{At the two-derivative level, after gauge fixing the theory will be closely related to cases where a global non-compact symmetry is broken perturbatively \cite{Burgess:2014, Csaki:2014}.}.

In order to obtain a \Poincare gravity theory, we employ the scale symmetry to impose $\chi^2 - \phi^2 = 6 \Mpl^2$, which has the solution
\bea
\frac{\chi}{\sqrt6 M_{pl}} =  \cosh \frac{\vf}{\sqrt{6\alpha} M_{pl}} \,, \quad \frac{\phi}{\sqrt6 M_{pl}}  = \sinh  \frac{\vf}{\sqrt{6\alpha} M_{pl}} \,.
\label{varphi}
\eea
in terms of a canonical scalar field $\vf$. Moreover, the vector field $b_\m$ can be eliminated by its field equation: for the above gauge choice it vanishes, while for other choices it is pure gauge. Most importantly, the scalar potential for $\vf$ reads
 \begin{align} \label{V-exp1}
  V = F(\tanh \frac{\vf}{\sqrt{6\alpha} M_{pl}}) \,,
 \end{align}
and is simply given by the function $F$ with the hyperbolic argument. The same behaviour was found for superconformal $\alpha$-attractors \cite{Kallosh:2013yoa}, which we will comment on in the conclusions.

For any finite value of $\alpha$, this implies that the scalar potential at large field values is flattened due to the hyperbolic functions (``inflation of moduli space'' instead of space-time). In this regime, the deviation from De Sitter will be governed by a single exponential drop-off term, and lead to a universal behaviour. Remarkably, this is virtually independent of the choice of the function $F$. In order to have the modes that we are observing inside this regime, the parameter $\alpha$ should be of order $\log(N)$ or smaller, where $N \sim 60$ is the number of e-folds since CMB horizon crossing. The resulting inflationary predictions are given by \cite{Kallosh:2013yoa}
 \begin{align}
   n_s = 1 - \frac2N \,, \quad r = \frac{12 \alpha}{N^2} \,, \label{predictions}
 \end{align}
up to higher-order corrections in $1/N$. For values of $\alpha \lesssim \log(N)$ where this approximation works, this yields an excellent fit to the measured Planck value of the spectral index, and moreover predict the tensor-to-scalar ratio of the order of a permille.
 
While this theory generically is invariant under \Poincare and scale transformations, it has a point of enhanced symmetry when $\alpha=1$. Exactly for this value the gauge field $b_\mu$ drops out of \eqref{2scalaractionS}, and can be rewritten in terms of conformal d'Alembertians for the two scalar fields. Therefore this point corresponds to a conformal theory, as pointed out in \cite{conformal-inflation}. This corresponds to the value $r= 0.003$ for $N=60$, similar to the Starobinsky theory \cite{Starobinsky:1980}. 

Another possibility for scale invariant theories is to take an $SO(2)$ symmetric combination instead: 
\begin{align}
\mathcal{L} = & \frac{1}{12} (\chi^2 + \phi^2) R(M) + \frac12 \a (\chi \Box^s \chi + \phi \Box^s \phi ) + \notag \\
&  -  \frac{1}{36} F\Big(\frac{\phi}{\chi}\Big) (\phi^2 + \chi^2)^2 \,,
\label{2scalaractionS2}
\end{align}
Note that this  is only possible for the scale invariant case, where the Einstein-Hilbert and scalar kinetic terms come with independent signs and coefficients, and does not have a conformal counterpart (the latter would require $\alpha=-1$ and leads to a ghost degree of freedom). The canonical inflaton is defined similar to \eqref{varphi} but now with trigonometric functions. One of the examples of this class is natural inflation \cite{Freese:1990}, which corresponds to the choice 
 \begin{align}
  F(x) = \frac{x^2}{1+x^2} \,, \quad V = \sin^2 \frac{\vf}{\sqrt{6\alpha} M_{pl}} \,.
 \end{align}
Other choices of $F$ correspond to different scalar potentials. In this case, there is no universal behavior for generic values of $\alpha$; the functional choice remains very relevant for the inflationary predictions, in contrast to the universal regime for the $SO(1,1)$ case.

\section{Higher derivative gravity}

In this section we will investigate the compatibility of extended symmetries with $f(R)$-theories of gravity, where the higher derivative terms induce the additional, inflationary degree of freedom. 

Our starting point will be the scale invariant action including an arbitrary function, $\mathcal{L} = e \, \chi^4 \, f(\Theta)$, where the argument is a linear combination of the two invariants:
\bea
 \Theta = \chi^{-2} R(M) - 6 \alpha \chi^{-3} \Box^s \chi \,.
\label{ScalefR2}
\eea
To gauge fix the dilatation symmetry, we set $\chi = \sqrt{6} \Mpl$. In order to convert this theory to a scalar-tensor theory, we take $\Theta$ to be an independent field and introduce a Lagrange multiplier $\varphi$ to enforce the relation \eqref{ScalefR2}. The field equations for $F$ and $b_\mu$ lead to
\bea
\frac{\partial f(\Theta)}{\partial \Theta} = e^{\sqrt{\frac{2}{3\a}} \frac{\vf}{M_{pl}}} \,, \quad b_\m = \frac{1}{\sqrt{6\a}} \frac{\partial_\m \vf}{M_{pl}} \,,
\label{feqsfr}
\eea
Hence we have traded the auxiliary field $\Theta$ and the gauge field $b_\mu$ for the canonically scalar field $\varphi$. Note that the latter is pure gauge, in accordance with having no propagating degrees of freedom. The scalar potential for the canonical inflation reads in Einstein frame \cite{Ozkan:2015}
\bea
  V= \frac12 M_{pl}^2 e^{-\sqrt{\frac{2}{3\a}} \frac{\vf}{M_{pl}}} [\Theta -    e^{-\sqrt{\frac{2}{3\a}} \frac{\vf}{M_{pl}}} f(\Theta) ]   \,,
\label{EquvAlpha}
\eea
which is fully determined by the choice of function $f(\Theta)$.

Turning to the conformal theory, the analogous construction is more restrictive: the only  invariant argument is
 \begin{align}
   \Theta = \chi^{-3} \Box^C \chi = - \frac16 \chi^{-2} R +  \chi^{-3} \Box \chi \,.
  \end{align}
Note that in this case one cannot add the square of covariant derivatives, as this term is not invariant under special conformal transformations.
The above is identical to the scale invariant argument, where $b_\mu$ drops out for the special choice $\alpha=1$. In this case the condition $b_\mu = 0$ arises as a gauge fixing condition for the special conformal transformations (in addition to $\chi = \sqrt{6} \Mpl$ for the dilatations). The two types of extensions of \Poincare symmetry therefore lead to similar results, with an $\alpha$-dependent scalar potential for the scale invariant case, while $\alpha=1$ in the conformal case. 

Specific choices for the function $f(R)$ deserve special mention. In the simplest example, when the function consists of a constant and a linear term, the theory has no additional propagating scalar degree of freedom: it is just general relativity with a cosmological constant. This is reflected in the solution \eqref{feqsfr} as these set $\varphi$ to a constant value. Secondly, the next-to-simplest combination of a linear and an (enhanced) quadratic term is the $\alpha$-generalization of Starobinsky inflation, with
\begin{align}
  V = V_0 ( 1 -  e^{-\sqrt{\frac{2}{3\a}} \frac{\vf}{M_{pl}}})^2 \,.
 \end{align}
For values of $\alpha$ up to order $\log(N)$, the predictions of this model are identical to \eqref{predictions}.

At this point we would like to stress that the same result follows from a much more general set of functions $f(R)$: any function that has an expansion of the form 
 \begin{align}
  f(R) = c_2 R^2 + c_1 R + c_0 1 + c_{-1} R^{-1} + \ldots \,,
 \end{align} 
will at large $R$ be dominated by the quadratic term that gives rise to a plateau at large and positive $\vf$ values, while the remaining terms yield exponentially suppressed drop-off terms. Only the leading of these, coming from the linear term in $f(R)$, will contribute to the scalar potential,
 \begin{align}
  V = \frac{1}{8 c_2} - \frac{c_1}{4 c_2} e^{-\sqrt{\frac{2}{3\a}}  \frac{\vf}{M_{pl}}} + \ldots \,.
  \end{align}
This determines the predictions \eqref{predictions} at this order; all other terms only come in at higher order in $1/N$.

Similarly, any function that has an expansion of the form 
 \begin{align}
  f(R) = c_2 R^2 + c_3 R^3 + c_4 R^4 + \ldots \,,
\end{align} 
will at small $R$ again be dominated by the quadratic term, which yields a plateau at large and negative $\vf$ values in this case:
 \begin{align}
  V = \frac{1}{8 c_2} - \frac{c_3}{16 c_2{}^3} e^{\sqrt{\frac{2}{3\a}}  \frac{\vf}{M_{pl}}} + \ldots \,.
  \end{align}
In this case the first of the exponentially suppressed drop-off terms is given by the quartic term, which yields identical predictions \eqref{predictions}.

Finally, the most general expansion \cite{Farakos:2013, Ferrara:2013b, Ozkan:2014, Broy:2014} 
 \begin{align}
  f(R) = \ldots + c_1 R + c_2 R^2 + c_3 R^3 + \ldots \,,
\end{align} 
 has an inflationary plateau with the same predictions, provided the quadratic term that leads to the scale-invariant plateau is sufficiently enhanced in order to be  long enough to generate $N$ e-folds.

\section{Conclusions}

Extending the \Poincare symmetry with a scaling symmetry has strong implications for inflation: very specific types of models arise naturally in set-ups with this extended symmetry. We have outlined this formalism for both two-derivative scalar-tensor theories as well as higher-derivative $f(R)$ theory. In contrast to \Poincare gravity, in the scale invariant and conformal cases the Einstein-Hilbert term comes with additional fields. Upon eliminating and gauge fixing these, this gives rise to scalar potentials \eqref{V-exp1} and \eqref{EquvAlpha} in terms of exponentially suppressed terms, whose argument is set by $\sqrt{2/3\alpha}$. Symmetry enhancement to the full conformal group occurs for $\alpha = 1$.

Interestingly, for a large range of values of the parameter $\alpha$, these lead to a single family of universality classes with the very simple predictions \eqref{predictions} indicated in Fig.~\ref{fig1}. In particular, the spectral index takes a definite value, which is perfectly compatible with Planck measurements. It is amusing to note that imposing scale invariance in the underlying theory implies a very specific deviation of a scale invariant spectrum of scalar perturbations, which would correspond to $n_s =1$. In addition, the tensor-to-scalar ratio has a simple expression and generically takes a permille value. 

The same predictions appeared earlier in the context of superconformal $\alpha$-attractors \cite{Kallosh:2013yoa}. In that construction, the $SO(1,1)$ symmetry was broken for any value of $\alpha \neq 1$; in the present framework, this would correspond to the inclusion of a d'Alembertian for the weight-zero combination $\phi/\chi$, thus breaking this global symmetry in the kinetic sector. In contrast, we retain both the scale symmetry and the $SO(1,1)$ symmetry for any value of $\alpha \neq 1$; the global symmetry is only broken by the profile functions $F(\phi / \chi) \neq 1$ or $f(R) \neq R^2$. It therefore appears that the full conformal symmetry and the crucial inflaton shift symmetry are only compatible when  $\alpha=1$.

Finally, it has been argued in \cite{Hertzberg:2014, Jackiw:2014} that the conformal symmetry in \cite{conformal-inflation} only provides a redundant description of a \Poincare invariant theory. This point is similar to the role of the superconformal symmetry in the construction of \Poincare supergravities \cite{Kaku:1978, Kallosh:2000, Freedman:2012} and whose relation to the Planck results was reviewed in \cite{Kallosh:2014}. However, while any inflationary model can be embedded in a scale invariant or conformal theory, for most models this will require more contrived constructions with a number of ingredients. In contrast, the universality classes studied in this letter  arise from the simplest possible construction  with either two scalars or an $f(R)$ theory. A case in point are $\alpha \neq 1$ models for a conformal theory: these would require a modification of the simple construction that we propose (e.g.~this would involve a weight-zero scalar field breaking the $SO(1,1)$ global symmetry). This underlines our central observation that
imposing scale or conformal invariance naturally leads to the promising set of inflationary models with specific non-minimal couplings and/or non-canonical kinetric terms. 



\section*{Acknowledgments}

We thank Gokhan Alkac, Souvik Banerjee, Renata Kallosh and Andrei Linde for  stimulating discussions. Moreover, we thank  Renata Kallosh and Andrei Linde for valuable comments on a draft of this letter including  the role of the superconformal symmetry for $\alpha \neq 1$.

\providecommand{\href}[2]{#2}\begingroup\raggedright\endgroup

\end{document}